 \documentclass[floatfix,fceqn]{elsart}
\usepackage{amsmath}
\usepackage{amssymb,graphicx,epstopdf,placeins,float}
\usepackage{diagbox}
\usepackage{pstricks}
\usepackage[numbers,sort&compress]{natbib}
\usepackage{hyperref}

 \newcommand{\beq}[1]{\begin{equation}\label{#1}}
 \newcommand{\eeq}{\end{equation}}
 \newcommand{\bea}[1]{\begin{eqnarray}\label{#1}}
 \newcommand{\eea}{\end{eqnarray}}
 \newcommand{\m}{{\rule[1.5pt]{2pt}{0.5pt}}}

\begin{document}

\begin{frontmatter}
\title{Exact Inner Metric and Microscopic State of AdS$_3$-Schwarzschld BHs}
\author[bjut]{Ding-fang Zeng}
\ead{dfzeng@bjut.edu.cn}
\address[bjut]{Institute of Theoretical Physics, Beijing University of Technology, China, Bejing 100124} 
\date{\today}
 \begin{abstract}
Through full solvability of 2+1 dimensional general relativity we derive out exact dynamic inner metric of collapsing stars with inhomogeneous initial mass distribution but joining with outside Anti-deSitt-Schwarzschild black holes smoothly. We prove analytically by standard quantum mechanics that the log-number of such solutions, or microscopic states of the system is proportional to the perimeter of the outside black holes. Key formulas for generalizing to 3+1D Schwarzschild black holes are also presented. Our result provides a bulk space viewpoint to questions on what the microscopic degrees of freedom are and who their carriers are in various holographic and/or asymptotic symmetry methods to black hole entropies. It may also shed light for singularity theorem and cosmic censorship related researches.
 \end{abstract}
 

\end{frontmatter}

\section{Introduction} 

In most works talking about the origin of Bekenstein-Hawking entropy, it is believed that  matters consisting of the black hole (BH) do not matter. What matters here is the gravitation field itself. A mostly cited reason for this belief is that, the matter's contribution can hardly be area law type, while holographic features of the gravitation theory make such contributions easy. For this reason, people invented many notion and methods such as asymptotic symmetry and soft hair \cite{Cardy1986, BrownHenneaux1986, sachs1962, BMS1962} et al to this question \cite{Carlip1995, Strominger1996, Strominger1997, Banados1998, Carlip2005, HPS2016}. Among these works, A. Strominger's calculation \cite{Strominger1997} of the entropy of 2+1 dimensional Banados-Teitelboim-Zanelli (BTZ\cite{BTZbh1992}) black holes is the most simply minded one: asymptotic symmetries of gravitations under consideration are nothing but those of two dimensional conformal field theory with appropriate central charges, micro-states of black holes in the former correspond to excitations of the latter with the same total energy in the Hilbert space and will exhibit perimeter --- area in 2+1 dimensions --- law as expected. By this working line, microscopic degrees of freedom of black holes are defined at infinites. Simple as it is \cite{Carlip1998}, it tells us very few on what they are describing and what their carriers are and contradicts even with some other approaches emphasizing excitations of the horizon.

However, logic possibilities that matters consisting the black hole being the main contributor of its Bekenstein-Hawking entropy are never excluded. Instead, string theory fuzzy ball pictures \cite{mathur2009,LuninMathur0202,mathur0502} strongly hint that black hole metrics with essential singularities are just overdone extrapolation of their outside geometry. Inside their horizon, essential singularities are resolved and matters/energy consisting them are regularly living there as strings, whose moving modes just form the basis of Bekenstein-Hawking entropy. On semi-classic levels, reference \cite{Stojkovic2008a,Stojkovic2015,Stojkovic2016} point out that to solve the information missing puzzle, physics of matter contents consisting of the black hole must be taken into account, and the central singularity of black holes must be resolved somehow at both classic and quantum levels \cite{Stojkovic2008b,Stojkovic2009,Stojkovic2014}. In a series of works \cite{dfzeng2017,dfzeng2018a,dfzeng2018b}, we throw away the hyper-physic notion such as extra-dimension and supersymmetry et al and considered the simple idea that matters consisting the black hole are not accreted on the central point statically but are oscillating around there under gravitations. This idea resolves the Schwarzschild singularity into a periodically density-divergent point thus giving matters consisting the black hole a chance to carry information without the singularity theorem\cite{SingularityPenrose,hawking1976,SingularityHawking}'s violation. In 4 and 5 dimensions, by looking matters inside the horizon of a dust ball as composites of many concentric spherical shells and quantize them canonically, we find that the number of distinguishable quantum states of them is finite and scales rather well as exponent of the horizon area with the equal mass Schwarzschild black holes. In the current work, we will use exact solvability of 2+1 dimensional general relativity to give these microscopic states' definition and enumeration an exact and more persuasive formulation.

In 2+1 dimensional asymptotically flat space-time, Newton's gravity has linear-inverse force $F\propto r^{-1}$, and linear-log potential $V\propto\log r$. As results, stars of any mass in this theory have infinite escaping speed, making the whole space-time behave as those in black holes do. In general relativities, Ricci tensor in 2+1 dimensional space-time has equal number of independent components as Riemann tensor does, so Ricci-flat means Riemann-flat too. As results \cite{Banados1996,Banados1997,Carlip1997,JBaez2007,Witten0706}, all particles with mass less than $(2G)^{-1}$ in the asymptotically flat spacetime will lead to locally flat geometry with conical singularities on the central point, while those with mass larger than $(2G)^{-1}$ always lead to spacetime with wrong signatured metric. However, in asymptotically AdS case, the background cosmological constant  provides a parabolic potential with minimal values zero. Adding effects of matters which are always trying to lower the potential everywhere, black holes with finite size can be formed,
\beq{}
ds^2=-hdt^2+h^{-1}dr^2+r^2d\phi^2
\label{AdSSbh}
\eeq
\beq{}
h=1+\frac{r^2}{\ell^2}-2GM
\label{hfunc}
\eeq
with the horizon radius given by $r_h=\sqrt{2GM-1}\ell$ and the corresponding Bekenstein-Hawking entropy given by the perimeter law  formula
\beq{}
S_\mathrm{BH}=\frac{k_{\scriptscriptstyle B}A^{(1)}_h}{4G}
=\frac{k_{\scriptscriptstyle B}2\pi r_h}{4G}
\xrightarrow{\frac{1}{2}\ll GM}\Big(\frac{k_{\scriptscriptstyle B}^2\pi^2M\ell^2}{2G}\Big)^\frac{1}{2}
\label{BHformula}
\eeq
The purpose of this work is to show, by general relativity and quantum mechanics, that this entropy arises from the way that $M$ can be distributed and moving inside the horizon, i.e. the way it can be looked as time dependent functions satisfying $M(t,r)_{r_h\leqslant r}=M_{tot}$, with the former understood as the mass $M$ inside radius $r$ ring at time $t$. For simplicity we will assume that $M$ consists of pure dust without pressure. Introducing pressure brings technical bothers but no key physics.

\section{Classic viewpoint} 
From general relativity aspects, we have two ways to know the form of $M(t,r)$. The first is through Einstein-equation $R_{\mu\nu}-\frac{1}{2}g_{\mu\nu}(R-2\Lambda)=8\pi G\rho u_{\mu}u_{\nu}+p(\cdots)$. However we must note here that, simply pumping a $t,r$ dependent $M$ into \eqref{AdSSbh}-\eqref{hfunc} will not yield proper geometry supported by physical fluid. We will use co-moving coordinate and write the metric in the matter occupation region as
\beq{}
ds^2=-d\tau^2\!+\!a^2(\tau)\big[\frac{d\varrho^2}{1\!+\!\varrho^2\ell^{\m2}\!-\!2GM(\varrho)}\!+\!\varrho^2d\phi^2\big]
\label{AdSmetricOscillation}
\eeq
\beq{}
a[\tau]=\cos[{\ell^{\m1}\tau}],~0\leqslant\varrho\leqslant\varrho_\mathrm{hor}~M[\varrho_{hor}]=M_\mathrm{tot}
\label{AdSmetricScaleFactor}
\eeq
Outside the horizon, $ds^2$ can be joined smoothly to \eqref{AdSSbh} by the method of standard textbooks \cite{GRweinberg}. We find that the above metric satisfies Einstein equation with $\Lambda=-\ell^{\m2}<0$ and 
\beq{}
\rho(\tau,\varrho)=\frac{M'\varrho^{-1}}{8\pi\,a[\tau]^2}
,~p=0
\label{massSuperpositionA}
\eeq
This expression tells us that singularities caused by gravitation collapse are only periodical but non-eternal. That is, if initially the mass distribution is regular, i.e. $M'\varrho^{-1}\neq\infty$, then such regularities will not be broken and kept static for ever.
Innocent as it looks, eqs.\eqref{AdSmetricOscillation}-\eqref{massSuperpositionA} provides us rather deep insight into microscopic state and inner structure of black holes. We can look $M(\varrho)$ as superposition of many concentric rings inside the horizon of an AdS$_3$-Schwarzshild black hole, each has radius $\varrho$ and carries mass $M' d\varrho$. As time passes by, all these rings will make equal-frequency $\ell^{\m1}$, equal phase but unequal amplitude --- $\varrho$-proportional--- oscillation under gravitations due to the background cosmological constant. Equal phase here is a very strong conclusion/condition, it means that matters inside the horizon make no shell-crossing motion. Otherwise, naked singularity \cite{Yodzis1973,Yodzis1974} will appear and the co-moving coordinate method will not apply at all. In real gravitation collapses, shell crossing is hardly to occur. Pressures arising from Pauli-exclusion principle will acts in such a way that initial speed of the matter collapsing becomes zero before gravitation force dominate all known physic expelling effects. 

The second way to know the form of $M(t,r)$ is to consider geodesic motion of test bodies co-moving with the dust volume element consisting and moving inside the horizon of the black hole. Looking the `hole' as many concentric rings, each of them has co-moving observers following orbit $\{t(\lambda),r(\lambda),\phi\m\mathrm{fixed}\}$\footnote{Non-trivial $\phi(\lambda)$ means orbits with nonzero angular momentum, the corresponding objects will not fall across the central point in any finite proper time. This means that classic black hole consisting of concentric shells each with nonzero angular momentum but adding up to zero are prohibited by the singularity theorem.} determined by
\beq{}
h_i\dot{t}^2-h^{-1}_i\dot{r}^2=1
,~h_i=1+\frac{r^2}{\ell^2}-2GM_i
\label{dssqEqOnedust}
\eeq
\beq{}
h_i\dot{t}=\gamma_i=\mathrm{const}
\Leftarrow\ddot{t}+\Gamma^{(i)t}_{tr}\dot{t}\dot{r}=0
\label{htdotEqconst}
\eeq
\beq{}
\Rightarrow\dot{r}^2-\gamma^2_i+h_i=0
\label{eomShellClassic}
\eeq 
where $M_i$ is the mass inside, including the $i$-th ring itself and $\Gamma^{(i)\sigma}_{\mu\nu}$ is the Christopher symbol associating with metrics $ds^2=-h_idt^2+h_i^{-1}dr^2+r^2d\phi^2$. Eq.\eqref{eomShellClassic} has oscillatory solution $r=r_\mathrm{imx}\cos(\omega_i\lambda+\varphi_i)$  with $r^2_\mathrm{imx}=2GM_i+\gamma^2_i-1$ and $t={\scriptstyle\displaystyle\int} h^{\m1}_i[r(\lambda)]\gamma_id\lambda$. When $r_\mathrm{imx}^2<(2GM_i-1)\ell^2$, $h_i$ will be negative and the corresponding $\lambda$ can be chosen to be pure imaginary. This does not affect the oscillatory behavior of $r(t)$. As results, the mass function $M(t,r)$ can be written down explicitly
\beq{}
M(t,r)=\sum_i m_i\Theta\{r-r_\mathrm{imx}\cos[\omega_i\lambda(t)+\varphi_i]\}
\label{massSuperpositionB}
\eeq
\beq{}
\sum_im_i=M_\mathrm{tot}, r^2_\mathrm{imx}=2GM_i+\gamma_i^2-1
\label{massAmplitudeB}
,~
\forall i,~\omega_i=\ell^{-1},~\mathrm{all}~\varphi_i~\mathrm{are~equal}
\eeq
where $\Theta[\cdots]$ is the usual Heaviside step function and we have labeled all the ring with their masses $m_i$, oscillation amplitudes $r_\mathrm{imx}$ and initial phases $\varphi_i$. Obviously, physics unveiled in this way are completely the same as those in the previous paragraph. Translating the function \eqref{massSuperpositionB} into the $M(\varrho)$ in \eqref{AdSmetricOscillation} is also a routine work except for some apparently singular coordinate transformation from $\{t,r\}$ to $\{\tau,\varrho\}$. By classic general relativity, the function form of both $M(\varrho)$ and $M(t,r)$ are uncountably infinite. However, at quantum levels, things are different and they form just the basis of microscopic state counted by the Bekenstein-Hawking entropy. 

\section{Quantum viewpoint} 
Taking an arbitrary mass ring from \eqref{massSuperpositionB} as an example, we can get its quantum description as follows:  introducing a wave function $\psi(r)$ to denote the probability amplitude it be measured at radius $r$, and operatorizing the radial momentum $m_i\dot{r}\equiv p_i$  as $p_i=i\hbar\partial_r$, then the quantum version of equation \eqref{eomShellClassic} will tell us
\beq{}
[(i\hbar m_i^{\m1}\partial_r)^2+1+\frac{r^2}{\ell^2}-2GM_i-\gamma^2_i]\psi(r)=0
\label{eqSchrodinger}
\eeq
In both this equation and its classic counterpart \eqref{eomShellClassic}, the radial coordinate $r$ is allowed to take values in the whole $0<r<\infty$ range. In classic equations, $\gamma^2_i<0$ will limit the mass shell to oscillate inside the horizon only; in quantum equation, $\gamma^2_i<0$ has no the power of forcing $\psi(r)$ to be zero outside the horizon uniformly . The square integrability of $\psi(r)$ in the whole $0<r<\infty$ range and $\gamma^2_i$'s being negative will assure that the maximal value of the $|\psi(r)|^2$ to occur inside or very near to the horizon.

After redefinitions, equation \eqref{eqSchrodinger}
\beq{}
r\rightarrow x,~\ell^{\m1}\rightarrow\omega, (2GM_i+\gamma^2_i-1)m_i\rightarrow 2E_i
\eeq
will become an eigen-state Schrodinger equation of harmonic oscillators
\beq{}
\big[-\frac{\hbar^2}{2m_i}\partial_x^2+\big(\frac{1}{2}m_i\omega^2x^2-E_i\big)\big]\psi(x)=0
\label{eqHarmOscillator}
\eeq
According to the standard quantum mechanic text books \cite{QMweinberg}, we know that due to the wave function's square integrability, energies of the mass shell are quantized,
\beq{}
\frac{E_i}{\hbar\omega}=\big(GM_i+\frac{\gamma^2_i-1}{2}\big)m_i\ell\hbar^{\m1}=n_i+\frac{1}{2}
\label{enQuantizCondition}
,~
n_i=0,1,2\cdots 
\eeq
with the corresponding wave function given by, here $N_i$ is the normalization constant
\beq{}
\psi^{E_i}_i[r]=N_ie^{-r^2\frac{m_i}{2\ell}}\mathrm{HermiteH}[\frac{E_i\ell}{\hbar}-\frac{1}{2},r\frac{m_i^\frac{1}{2}}{\ell^\frac{1}{2}}]
\label{wfHarmOscillator}
\eeq
Comparing \eqref{enQuantizCondition} and \eqref{massAmplitudeB}, we easily see that in classic theories the mass/energy of a composite ring is determined by its oscillation amplitude, while in quantum theories it has no direct relation with the amplitude but are integer multiples of the oscillation frequency.  Nevetheless, \eqref{enQuantizCondition} tells us that $E_i$ has two origins, the first is the gravitational part $GM_im_i$, while the second is the kinetic contribution $\frac{\gamma_i^2-1}{2}m_i$.  For all shells released inside the horizon $\gamma_i^2<-1$,  this two term contribute contrarily to $E_i$.

The eigen energy and wave function quotient above tell us that the ring under consideration can only be at a series of quantum state marked by $n_i$, while the disk or black hole as a whole can only be at states featured by direct products of its composite rings' wave function. The probability that we find mass $M$ inside the radius $r$ circle at time $t$ is
\beq{}
P[M,t,r]\propto\int_0^r\left|\sum_i^{\sum E_i=M\ell/G}e^{\frac{iE_it}{\hbar}}\psi_i^{E_i}[\hat{r}]\right|^22\pi\hat{r}d\hat{r}
\eeq
This is our quantum version of classic mass function \eqref{massSuperpositionB}. However, we must be cautioned with subtleties involved in the physic interpretation of $t$ and $E_i$. If we take the viewpoint that \eqref{eqHarmOscillator}-\eqref{enQuantizCondition} is a flat space quantum oscillator, then they have the meaning of time and energy definitely. However, such a viewpoint is valid only locally according to the equivalent principle. Away from any pre-specified reference point, we must consider the redshift effect on them. 

Consider the relation between $m_i$ and $M_\mathrm{tot}$, when we neglect gravitation and relativity effects, $\sum m_i=M_\mathrm{tot}$. However, when such effects are considered, the relation becomes $\sum E_i\frac{G}{\ell}=M_\mathrm{tot}$. The factor $\frac{G}{\ell}$ on the left hand side is due to the redshift  because $E_i$ from \eqref{eqHarmOscillator}-\eqref{enQuantizCondition} can be understood energy only locally on the origin, while $M_\mathrm{tot}$ is an invariant in AdS with time warps $\frac{r^2}{\ell^2}$. Considering this fact, energies of the composite ring and the mass of the whole disk or black hole should be written as
\beq{}
\sum({n_i}+\frac{1}{2})\hbar\omega=M_\mathrm{tot}\frac{\ell}{G}
\label{quantumMassRelation}
\eeq
From this relation and the famous partition number formula of Ramanujan \cite{Ramanujan1918}, we know in the large $M\ell^2/{\hbar}G$ limit, the way of building the whole disk or black hole has the following number of possibilities
\beq{}
W=\exp\{2\pi\sqrt{\frac{1}{6}\frac{M_\mathrm{tot}\ell^2}{\hbar G}}\}
\eeq
This is nothing but the origin of Bekenstein-Hawking entropy given in the perimeter law formula \eqref{BHformula} except for a pure numeric factor of $\big(\frac{4}{3}\big)^{_{1/2}}$, i.e.  $k_{\scriptscriptstyle B}\log{W}=\big(\frac{4}{3}\big)^{_{1/2}} S_\mathrm{BH}$, which arises from our imprecise estimation of the redshift factor $\frac{\ell}{G}$ in \eqref{quantumMassRelation}. In approaches basing on asymptotic symmetries, this factor is interpreted as central charges of the corresponding conformal field theory \cite{Carlip1998}. Here, we see that it may have a different mechanism of origin. No matter where its origin is from, and even no matter it originates or not, we emphasize here that the key feature of Bekenstein-Hawking entropy $S_\mathrm{BH}\propto M^\frac{1}{2}$ follows simply from the radial distribution modes counting of matters consisting and moving inside the horizon of an AdS-Schwarzschild black hole.

On this microscopic origin of Bekenstein Hawking entropy, the most direct question is, why do we neglect the black hole consisting particles' random motion degrees of freedom. The answer is related with the singularity theorem, according to which all matters falling into the horizon or consisting of a Schwarzschild black hole are to pass through the central point at finite proper time. So they have zero angular momentum and no non-radial motion is allowed. Except singularity theorem, another reason denying the random motion's contribution to $S_\mathrm{BH}$ is that, in AdS$_3$-Schwarzschild black holes $S_\mathrm{BH}=2\pi r_h\propto M^\frac{1}{2}<N$, while in 3+1-Schwarzschild case $S_\mathrm{BH}=4\pi r_h^2\propto M^2>N$, with $N$ being the number of particles in the system and $S_\mathrm{rand}\propto N$ by general statistic mechanics. Obviously the scaling law produced by this contribution to $S_\mathrm{BH}$ is contradict to each other in 2+1-AdS-Schwz and 3+1-Schwz black holes.

\section{3+1D generalizations} 

In 3+1 Schwarzschild black holes, completely parallel with \eqref{AdSmetricOscillation} \eqref{AdSmetricScaleFactor} \eqref{massSuperpositionA} and \eqref{eqHarmOscillator} \eqref{enQuantizCondition} \eqref{wfHarmOscillator}, we have
\beq{}
ds^2=-d\tau^2\!+\!\frac{\big[1\!-\!\big(\frac{2GM}{\varrho^3}\big)\!^\frac{1}{2}\!\frac{M'\!\varrho}{2M}\tau\big]^2\!d\varrho^2}{a[t,\varrho]}+a[\tau,\varrho]^2\varrho^2d\Omega^2_2
\label{genOSmetric}
\eeq
\beq{}
a[\tau,\varrho]\!=\!\big[1-\frac{3}{2}\big(\frac{2GM[\varrho]}{\varrho^3}\big)\!^\frac{1}{2}\tau\big]^\frac{2}{3}
,~M[\varrho_\mathrm{hor}\!\leqslant\varrho]=M_\mathrm{tot}
\label{genOSscalefactor}
\eeq
\beq{}
\rho[\tau,\varrho]=\frac{M'[\varrho]/8\pi\varrho^2}{a^\frac{3}{2}
\!+\!\frac{3GM'\tau^2}{4\varrho^2}
\!-\!\big(\frac{GM}{\varrho^3}\big)\!^\frac{1}{2}\frac{M'\varrho\tau}{2M}}
,~p=0
\label{genOSdustDistribution}
\eeq
\beq{}
\big[-\frac{\hbar^2}{2m_i}\partial_x^2-\frac{GM_im_i}{x}-E_i\big]\psi(x)=0,~
\eeq
\beq{}
E_i=\frac{\gamma^2_i\!-\!1}{2}m_i=-\frac{(GM_im_i)^2m_i}{n_i^2\hbar^2}
,~n_i=1,2,\cdots
\eeq
\beq{}
\psi^{E_i}_i[r]=N_ie^{-\hat{r}}\hat{r}\mathrm{LaguerreL}[n_i\!-\!1,1,2\hat{r}],
\hat{r}\equiv\frac{m_ir}{1-\gamma_i^2}
\label{wfHydrogenAtom}
\eeq
In contrast with AdS2+1 black holes, where the cosmological constant induced AdS-potential well provides the condition for matter oscillations inside the horizon, in the 3+1 Schwarzschild black holes, it is self-gravitation attraction that provides the driving force for matter oscillations inside the horizon. Due to the quantum forward scattering effects, matters falling towards the central point of the black hole do not form static singularities, but over-forward walking and crossing each other there. This forward scattering or over-forward walking is the key for central singularities' resolution and oscillations inside the horizon.

Also different from the AdS2+1 cases, in 3+1D Schwarzschild black holes the mass function is oscillating in a $\big(1-\frac{\tau}{T/4}\big)\!^\frac{2}{3}$ pattern\footnote{Formulas \eqref{genOSmetric} \eqref{genOSscalefactor} \eqref{genOSdustDistribution} are generalizations of the Opernheimer-Snyder collapsing star \cite{oppenheimer1939,MTWbook}. Our formulas describe dynamic collapsing stars with inhomogeneous but spherical symmetric initial mass distribution whose outside geometry joins to the Schwarzschild metric smoothly. Without pressure, we cannot set the initial speed of the system to be zero mathematically. In real collapsing stars, pressures originating from the Pauli exclusion principle can support a zero initial speed collapsing. $\big(1-\frac{\tau}{T/4}\big)\!^\frac{2}{3}$ is the behavior of the oscillation across the central point} instead of the harmonic $\cos[\ell^{-1}\tau]$ one. However, this difference is completely irrelevant for the Schwarzschild singularity's resolving at both classic and quantum levels. Also the same as AdS$_3$-Schwarzschild holes, $\psi_i^{E_i}[r]_{r_\mathrm{hor}<r}\neq0$ provides us a rather intuitive fuzzball picture for black holes. That is, we have always nonzero --- although very small --- probabilities to find the composite shells outside the horizon. Requiring all compositing shells' energy add up to $-M_\mathrm{tot}$ yields similar equality like \eqref{quantumMassRelation}, $\sum_i\frac{G^2M_i^2m_i^3}{n_i^2\hbar^2}=M_\mathrm{tot}$. But in this case we have no adoptable Ramanujan formula to derive the area law entropy analytically. Nevertheless, in refs. \cite{dfzeng2018a} by listing out all shell-partitioning method which lead to distinguishable direct product quantum state in small black holes, we provide substantial evidence that the area law entropy indeed follows from this radial mass' moving modes counting. Basing on this picture, we  provide in \cite{dfzeng2018b} even a detailed calculation of spontaneous radiation rates of the direct product quantum system to uncover where the information is going during hawking radiations. 

\section{Conclusion and Discussion} 

Come back to our beginning talk about the origin of Bekenstein-Hawking entropy. Both our exact solution and analytical proof in the main text support that the radial collective motion modes of matters consisting of an AdS$_3$-Schwarzschild black hole is the main contributor to the area (perimeter in 2+1D spacetime) law feature of Bekenstein-Hawking formulas. The popular belief that such object's contribution should be proportional to the volume (area in 2+1D) of the system is a doctrine completely.  Classic particles inside the horizon carrying non-zero angular momentum do not go across the central point in any finite proper time, so is prohibited by  singularity theorems. They makes no contribution to $S_\mathrm{BH}$. 

Comparing two set of equal mass binary black hole system fixed on $z$-axis and separated by $2r_h=4Gm$, the former consists of two point like singular black holes while the latter consists of two oscillatory uniform density regular ones, their quadrupoles can be calculated easily
\beq{}
\begin{array}{c}D^p_{zz}=2mr_h^2,~D^p_{ij(\neq zz)}=0
\vspace{3mm}\\
D^b_{zz}=\frac{12}{5}mr_h^2,~D^b_{xx}=D^b_{yy}=\frac{2}{5}mr_h^2,~D^{b}_{ij(i\neq j)}=0
\end{array}
\eeq
When this two set of binary system rotate around their own central vertical line and radiate gravitational waves \cite{gw150914, gw170817a, gwScalarSoliton}, the wave will have unequal amplitude and can be measured by us. So our proposition in this and previous works \cite{dfzeng2017,dfzeng2018a,dfzeng2018b} that matters consisting black holes are regularly living and oscillating inside the horizon instead of singularly accumulating on the central point is dis-verifiable through gravitational wave observations from the binary black hole merging events. This disprovability  or allowing inside-horizon structure be measurable to outside observers does not violate causality, because when quantum effects are considered, the black hole horizon is highly blurred. This is easy to understand from the wave function \eqref{wfHarmOscillator} and \eqref{wfHydrogenAtom}'s nonzero tail outside the horizon.

From pure theoretic aspects, our work provides a bulk space answer to questions such as what the microscopic degrees of freedom are and who their carriers are in various holographic and/or asymptotic symmetry methods to black hole entropies. Superficially looking, our oscillatory shell explanation and Strominger-Carlip's near-horizon or asymptotic symmetry \cite{Strominger1997,Carlip1998} explanation for black hole's microstate contradict each other. However, considering the fact that all microstate black holes in our explanation have equal outside horizon symmetry, and the precise map between the classic inside-horizon metric of AdS2+1 black holes and quantum states of the appropriate near-horizon or boundary conformal field system are not clearly known, this contradiction may not necessarily be the case. To figure out this map in detail is a valuable direction for future exploration. Our work may also shed light for singularity theorem and cosmic censorship related researches.

\section*{Acknowledgements}
This work is supported by NSFC grant no. 11875082.


\begin{thebibliography}{99}

\bibitem{Cardy1986}
H. Bloete, J. Cardy, M. Nightingale 
``Conformal Invariance, the Central Charge, and Universal Finite Size Amplitudes at Criticality'',
{\em Phys. Rev. Lett.} {\bf56} (1986) 742.

\bibitem{BrownHenneaux1986}
J. Brown, M. Henneaux,
``Central Charges in the Canonical Realization of Asymptotic Symmetries: An Example from Three-Dimensional Gravity'',
{\em Commun. Math. Phys.} {\bf104} (1986) 207-226.

\bibitem{sachs1962}
R. Sachs, ``Asymptotic Symmetries in Gravitational Theory'',
{\em Phys. Rev.} {\bf128} (1962) 2851.

\bibitem{BMS1962}
H. Bondi, M. van der Burg, A. Metzner,
``Gravitational Waves in Generral Relativity, 7: Waves from axisymmetric isolated systems'',
{\em Proc. Roy. Soc. Lond.} {\bf A269} (1962) 21

\bibitem{Carlip1995}
S. Carlip,
``The Statistical Mechanics of the (2+1)-Dimensional Black Hole'',
{\em Phys. Rev.} {\bf D51} (1995) 632,
\href{https://arxiv.org/abs/gr-qc/9409052}{gr-qc/9409052}

\bibitem{Strominger1996}
A. Strominger, C. Vafa,
``Microscopic Origin of the Bekenstein-Hawking Entropy''
{\em Phys. Lett} {\bf B379} (1996) 99,
\href{https://arxiv.org/abs/hep-th/9601029}{arXiv: 9601029}.

\bibitem{Strominger1997}
A. Strominger,
``Black Hole Entropy from Near-Horizon Microstates'',
{\em JHEP} {\bf 02} (1998) 009,
\href{https://arxiv.org/abs/hep-th/9712251}{hep-th9712251}.

\bibitem{Banados1998}
M. Banados, T. Brotz, M. Ortiz,
``Boundary dynamics and the statistical mechanics of the 2+1 dimensional black hole'',
{\em Nucl. Phys.} {\bf B545} (1998) 340,
\href{https://arxiv.org/abs/hep-th/9802076}{hep-th/9802076}.

\bibitem{Carlip2005}
S. Carlip,
``Conformal Field Theory, (2+1)-Dimensional Gravity, and the BTZ Black Hole'',
{\em Class. Quant. Grav.} {\bf22} (2005) R85-R124,
\href{https://arxiv.org/abs/gr-qc/0503022}{gr-qc/0503022}.

\bibitem{HPS2016}
``Soft Hair on Black Holes''
S. Hawking, M. Perry, A. Strominger,
{\it Phys. Rev. Lett.} {\bf116} (2016) 231301,
\href{http://arxiv.org/abs/1601.00921}{arXiv:1601.00921}

\bibitem{BTZbh1992}
M. Ba$\tilde{n}$ados, C. Teitelboim, J. Zanelli,
``The Black Hole in Three Dimensional Space Time'',
{\em Phys. Rev. Lett.} {\bf69} (1992) 1849,
\href{https://arxiv.org/abs/hep-th/9204099}{hep-th/9204099}.

\bibitem{Carlip1998}
S. Carlip,
``What we don't know about BTZ black hole entropy'',
{\em Class. Quantum Grav.} {\bf 15} (1998) 3609,
\href{http://arxiv.org/abs/hep-th/9806026}{ePrint: hep-th/9806026}.

\bibitem{mathur2009}
S. D. Mathur,
``The information paradox: A pedagogical introduction'',
{\em 	Class.Quant.Grav.} {\bf 26:} 224001(2009),
\href{https://arxiv.org/abs/0909.1038}{arXiv: 0909.1038}

\bibitem{LuninMathur0202}
O. Lunin, S. D. Mathur,
``Statistical interpretation of Bekenstein entropy for systems with a stretched horizon'',
{\em Phys. Rev. Lett.} {\bf 88} (2002) 211303,
\href{https://arxiv.org/abs/hep-th/0202072}{arXiv: hep-th/0202072}.

\bibitem{mathur0502}
S. D. Mathur,
``The Fuzzball proposal for black holes: An Elementary review'',
{\it Fortsch.Phys.} {\bf53} (2005) 793-827,
\href{http://arxiv.org/abs/hep-th/0502050}{arXiv: 0502050}.

\bibitem{Stojkovic2008a}
T. Vachaspati, D. Stojkovic;
``Quantum Radiation from Quantum Gravitational Collapse'',
{\em Phys. Lett.} {\bf B663} (2008) 107-110,
\href{https://arxiv.org/abs/gr-qc/0701096}{ePrint: gr-qc/0701096}.

\bibitem{Stojkovic2015}
A. Saini, D. Stojkovic,
``Radiation from a collapsing object is manifestly unitary'',
{\it Phys. Rev. Lett.} {\bf114} (2015), 111301;
\href{https://128.84.21.199/abs/1503.01487v3}{arXiv: 1503.01487}.

\bibitem{Stojkovic2016}
J. Hutchinson, D. Stojkovic,
``Icezones instead of firewalls: extended entanglement beyond the event horizon and unitary evaporation of a black hole'',
{\em Class. Quant. Grav.} {\bf33} (2016) no.13, 135006;
\href{http://arxiv.org/abs/arXiv:1307.5861}{arXiv: 1307.5861}.

\bibitem{Stojkovic2008b}
E. Greenwood, D. Stojkovic,
``Quantum gravitational collapse: Non-singularity and non-locality'',
{\em JHEP} {\bf 0806} (2008) 042;
\href{http://arxiv.org/abs/arXiv:0802.4087}{arXiv: 0802.4087}.

\bibitem{Stojkovic2009}
J. Wang, E. Greenwood, D. Stojkovic,
``Schrodinger formalism, black hole horizons and singularity behavior'',
{\em Phys. Rev.} {\bf D80} (2009) 124027;
\href{http://arxiv.org/abs/arXiv:0906.3250}{arXiv: 0906.3250}.

\bibitem{Stojkovic2014}
A. Saini, D. Stojkovic,
``Nonlocal (but also nonsingular) physics at the last stages of gravitational collapse'',
{\em Phys. Rev.} {\bf D89} (2014) no.4, 044003;
\href{http://arxiv.org/abs/arXiv:1401.6182}{arXiv: 1401.6182}.

\bibitem{dfzeng2017}
Ding-fang Zeng, 
``Resolving the Schwarzschild singularity in both classic and quantum gravities'',
{\em Nucl. Phys.} {\bf B917} 178-192,
\href{http://arxiv.org/abs/arXiv:1606.06178}{arXiv: 1606.06178}.

\bibitem{dfzeng2018a}
Ding-fang Zeng, 
``Schwarzschild Fuzzball and Explicitly Unitary Hawking Radiations'',
{\em Nucl. Phys.}{\bf B930} (2018) 533-544,
\href{http://arxiv.org/abs/arXiv:1802.00675}{arXiv: 1802.00675}.

\bibitem{dfzeng2018b}
Ding-fang Zeng, 
``Information missing puzzle, where is hawking's error?'',
{\em Nucl. Phys.}{\bf B} in production,
\href{https://doi.org/10.1016/j.nuclphysb.2019.02.023}{https://doi.org/10.1016/j.nuclphysb.2019.02.023},
\href{http://arxiv.org/abs/arXiv:1804.06726}{arXiv: 1804.06726}.

\bibitem{SingularityPenrose}
R. Penrose, 
"Gravitational collapse and space-time singularities", 
{\it Phys. Rev. Lett.} {\bf14} (1964) 57.

\bibitem{hawking1976}
S. Hawking,
``Breakdown of Predictability in Gravitational Collapse'',
{\em Phys. Rev. } {\bf D14, } 2460 (1976)

\bibitem{SingularityHawking}
S. Hawking, G. F. Ellis,
``The Large Scale Structure of Space Time''
Cambridge: Cambridge University Press. ISBN 0-521-09906-4.

\bibitem{Banados1996}
M. Banados,
``Global Charges in Chern-Simons theory and the 2+1 black hole'',
{\em Phys. Rev.} {\bf D52} 1996 (5816),
\href{https://arxiv.org/abs/hep-th/9405171}{hep-th/9405171}.

\bibitem{Banados1997}
M. Banados, A. Gomberoff,
``Black Hole Entropy in the Chern-Simons Formulation of 2+1 Gravity''
{\em Phys. Rev} {\bf D55} (1997) 6162,
\href{https://arxiv.org/abs/gr-qc/9611044}{gr-qc/9611044}.

\bibitem{Carlip1997}
S. Carlip,
``The Statistical Mechanics of the Three-Dimensional Euclidean Black Hole'',
{\em Phys. Rev.} {\bf D55} (1997) 878,
\href{https://arxiv.org/abs/gr-qc/9606043}{gr-qc/9606043}.

\bibitem{JBaez2007}
J. Baez, D. Wise, A. Crans,
``Exotic Statistics for Strings in 4d BF Theory'',
{\em Adv. Theor. Math. Phys.} {\bf11} (2007) 707,
\href{https://arxiv.org/abs/gr-qc/0603085}{gr-qc/0603085}.

\bibitem{Witten0706}
E. Witten,
``Three Dimensional Gravity Reconsidered'',
\href{http://arxiv.org/abs/arXiv:0706.3359}{arXiv: 0706.3359}

\bibitem{QMweinberg}
S. Weinberg,
``Lectures on Quantum Mechanics'', chapter 2,
Cambridge: Cambridge University Press. ISBN 978-1-107-02872-2.

\bibitem{Yodzis1973}
P. Yodzis, H.J. Seifert and H. Müller zum Hagen,
``On the occurrence of naked singularities in general relativity'',
{\it Commun. Math. Phys.} {\bf34} (1973), 135.

\bibitem{Yodzis1974}
P. Yodzis, H.J. Seifert and H. Müller zum Hagen,
``On the occurrence of naked singularities in general relativity. II'',
{\it Commun. Math. Phys.} {\bf37} (1974), 29.

\bibitem{GRweinberg}
S. Weinberg, 
“Gravitation and Cosmology: Principles and Applications of the General Theory of Relativity”, chapter 11, 
Wiley and Sons, New York, 1972, ISBN-13: 978-0471925675

\bibitem{Ramanujan1918}
S. Ramanujan and G. Hardy,
{\em Proc. London Math. Soc.(ser.2)} {\bf17} 75, 1918,
reprinted in G. Hardy et al (ed), 
{\em Collected papers of Srinivase Ramanujan} (New York: Chelsea).

\bibitem{oppenheimer1939} J.R. Oppenheimer and H. Snyder, ``On continued gravitational contraction'',
{\it Phys. Rev.} {\bf56} (1939) 455.

\bibitem{MTWbook} C. Misner, K. Thorne, J. Wheeler, ``Gravitation'',
{\it ``Gravitation''} {\bf\S32.3,32.4,32.6}, W. H. Freeman and Company Version 1973.

\bibitem{gw150914}
The LIGO Scientific Collaboration, the Virgo Collaboration,
``Observation of Gravitational Waves from a Binary Black Hole Merger'',
{\it Phys. Rev. Lett.} {\bf116} (2016) 061102,
\href{https://arxiv.org/abs/1602.03837}{arXiv:1602.03837}.

\bibitem{gw170817a}
Abbott, B. P.  et al. (LIGO Scientific Collaboration \& Virgo Collaboration),
"GW170817: Observation of Gravitational Waves from a Binary Neutron Star Inspiral",
{\it Phys. Rev. Lett.} {\bf119} (2017) 161101,
\href{https://arxiv.org/abs/1710.05832}{arXiv:1710.05832 }

\bibitem{gwScalarSoliton}
T. Helfer, E. A. Lim, M. A. Garcia, M. A. Amin,
``Gravitational Wave Emission from Collisions of Compact Scalar Solitons'',
\href{https://arxiv.org/abs/1802.06733}{arXiv: 1802.06733}

\end{thebibliography}
\end{document}